\documentclass[%
 reprint,
 prd,twocolumn,
 preprintnumbers,nofootinbib,
 amsmath,amssymb,
 longbibliography,floatfix]{revtex4-2}
\usepackage[utf8]{inputenc}
\usepackage{graphicx}
\usepackage{bm}
\usepackage{hyperref}
\usepackage{orcidlink}

\begin{document}


\title{Dispersion Relations of Collective Flavor Excitations in Dense Neutrino Media}
\newcommand{\UNM}{Department of Physics and Astronomy, University of New Mexico, Albuquerque, New Mexico 87131, USA}

\author{Anson Kost\,\orcidlink{0009-0008-0327-5057}}
\author{Huaiyu Duan\,\orcidlink{0000-0001-6708-3048}}

\affiliation{\UNM}

\date{\today}

\begin{abstract}
We apply the Landau damping theory to collective flavor excitations in dense neutrino media. We demonstrate explicitly that the (meta)stable modes of the collective flavor excitations in a physical dense neutrino gas are indeed weakly damped. We show that, in a neutrino medium with a continuous energy distribution, there can exist infinitely many dispersion relation branches of collective flavor excitations that extend to large wave numbers. In a neutrino gas with one or more zero crossings in the distribution of the difference of the (neutrino) lepton numbers (DLNs), the least damped resonant branches and the metastable branches can suffer from inverse Landau damping at certain wave numbers and thus become unstable. We also introduce the concept of convolved DLN distributions, which can be used to understand the physical origin of the slow and fast flavor instabilities in dense neutrino media in core-collapse supernovae and neutron star mergers.
\end{abstract}

\maketitle

\section{Introduction}
\label{sec:intro}

Neutrinos of different flavors can transform into each other in vacuum or in matter, a quantum phenomenon known as neutrino oscillations (see, e.g., \cite{ParticleDataGroup:2024cfk} for a review). This phenomenon is particularly interesting in core-collapse supernovae and neutron star mergers, where the extremely dense neutrino media can experience flavor oscillations collectively (see, e.g., \cite{Duan:2010bg,Chakraborty:2016yeg,Tamborra:2020cul,Volpe:2023met,Johns:2025mlm} for some reviews). Collective neutrino oscillations can help shape the physical and chemical evolution of these astrophysical environments and leave imprints on the neutrino signals from these events that may be detected in the future.

An important question for collective neutrino oscillations is under what conditions they will occur. This question is usually addressed by analyzing the linear stability of the equations of motion (EoM) with small flavor coherences \cite{Banerjee:2011fj}. Let
\begin{align}
    \rho_{\mathbf{p}}(t, \mathbf{r}) = \begin{pmatrix}
        f_{\mathbf{p}}^{\nu_e} & \psi_{\mathbf{p}}^*/2 \\
        \psi_{\mathbf{p}}/2 & f_{\mathbf{p}}^{\nu_\mu}
    \end{pmatrix}
\end{align}
be the (two-flavor) flavor density matrix of the neutrino medium for the neutrinos of momentum $\mathbf{p}$ at time $t$ and position $\mathbf{r}$ \cite{Sigl:1993ctk}, where $f^{\nu_{a}}_\mathbf{p}$ and $\psi_\mathbf{p}$ are the occupation number in flavor $a$ ($a=e,\mu$) and the flavor coherence, respectively, of the neutrinos. A similar quantity $\bar\rho_\mathbf{p}$ can be defined for the antineutrinos.  An eigenmode of the linearized EoM takes the form
\begin{align}
    \psi_{\mathbf{p}}(t, \mathbf{r}), \bar\psi_{\mathbf{p}}(t, \mathbf{r}) \propto e^{\mathrm{i}[\mathbf{K}\cdot\mathbf{r} - (\Omega + \mathrm{i}\Gamma) t]},
\end{align}
where $\mathbf{K}$, $\Omega$, and $\Gamma$ are the wave vector, the frequency, and the growth rate of the eigenmode, respectively. A flavor instability occurs if $\Gamma > 0$, which indicates that the flavor coherence grows exponentially with time. The eigenmodes of the linearized EoM can be obtained from a self-consistency equation for the total flavor coherence \cite{Izaguirre:2016gsx}
\begin{align}
    \Psi^\alpha = \int v^\alpha (\psi_\mathbf{p} - \bar{\psi}_\mathbf{p})\,\frac{\mathrm{d}^3p}{(2\pi)^3},
\end{align}
where $(v^0, \mathbf{v}) = (1, \mathbf{p}/|\mathbf{p}|)$ is the four-velocity of an ultra-relativistic neutrino of momentum $\mathbf{p}$.

The flavor instabilities in a neutrino medium are known to be related to the distribution of the difference of the (neutrino) lepton numbers (DLN) $g(\omega, \mathbf{v})$ \cite{Dasgupta:2009mg}, which is proportional to $f_{\mathbf{p}}^{\nu_e} - f_{\mathbf{p}}^{\nu_\mu}$ for neutrinos and $f_{\mathbf{p}}^{\bar\nu_\mu} - f_{\mathbf{p}}^{\bar\nu_e}$ for antineutrinos. In this paper, we adopt the convention of the neutrino flavor isospin in which the vacuum frequencies $\omega$ of the neutrinos and antineutrinos take opposite signs \cite{Duan:2005cp}. The early literature focused on the systems that are \emph{equivalent} to homogeneous and isotropic neutrino gases%
\footnote{In the stationary ``bulb model'' of supernova neutrinos \cite{Duan:2006an}, the radial direction plays the role of ``time'', and the homogeneity and isotropy are replaced by the spherical symmetry about the center of the ``neutrino bulb'' and the axial symmetry about the radial direction, respectively.}
where the (effective) DLN distribution $g(\omega, \mathbf{v}) = g(\omega)$ is independent of the neutrino velocity $\mathbf{v}$. It was observed that a (zero) crossing in $g(\omega)$ with a negative (positive) slope is related to the flavor instability of the isotropy-preserving (isotropy-breaking) modes \cite{Dasgupta:2009mg, Raffelt:2013rqa, Duan:2013kba}. Most of the recent literature instead focuses on the pure fast modes which are determined by the DLN angular distribution
\begin{align}
    \hat{g}(\mathbf{v}) = \int_{-\infty}^\infty g(\omega, \mathbf{v})\,\mathrm{d}\omega.
\end{align}
A crossing in $\hat{g}(\mathbf{v})$ is both necessary and sufficient for the occurrence of flavor instabilities in the pure fast limit \cite{Morinaga:2021vmc}. It has been further shown that a crossing in $g(\omega, \mathbf{v})$ is a necessary condition for any flavor instability \cite{Dasgupta:2021gfs}. 

A new perspective on flavor instabilities has emerged in the past couple of years based on the similarity between the quantum kinetic equation (QKE) for a dense neutrino medium and the Vlasov equation for a collisionless plasma \cite{Fiorillo:2024bzm,Fiorillo:2024uki,Fiorillo:2024pns,Fiorillo:2025ank,Fiorillo:2025zio}. Instead of being treated as an auxiliary quantity to obtain the eigenmodes of the linearized EoM, $\Psi^\alpha$ takes center stage and takes on the physical meaning of the collective flavor excitation of the neutrino medium.
A growing flavor excitation is identified with a flavor instability in the eigenmode approach. It occurs when the collective mode is in resonance with a ``flipped'' section of the DLN distribution that has a sign opposite to that of the total DLN. A decaying flavor excitation, on the other hand, does not necessarily imply the decay of the flavor coherences $\psi_{\mathbf{p}}$ and $\bar\psi_\mathbf{p}$ of the neutrino field, but rather is a result of the phase mixing of the non-collective eigenmodes \cite{Capozzi:2019lso,Fiorillo:2024bzm}. It occurs when the collective mode is in resonance with a part of the DLN distribution that has the same sign as the total DLN. This phenomenon is similar to the Landau damping of plasma waves \cite{Landau:1946jzi}.

The quantized version of the flavor excitation is called the \emph{flavomon} \cite{Fiorillo:2025npi,Fiorillo:2025gkw,Fiorillo:2025kko}. A flavomon carries momentum $\mathbf{K}$, energy $\Omega$, $+1$ unit of the $\mu$ lepton number and $-1$ unit of the $e$ lepton number, and has a mean lifetime $(2|\Gamma|)^{-1}$ if $\Gamma< 0$. It can be absorbed or emitted through the processes 
\begin{align}
    \nu_\mu \rightleftharpoons \nu_e + \Psi
    \qquad\text{and}\qquad
    \bar\nu_e \rightleftharpoons \bar\nu_\mu + \Psi.
    \label{eq:processes}
\end{align}
Because the self-interaction strength $\mu$ of the neutrino medium is much smaller than the neutrino kinetic energy $|\mathbf{p}|$, the neutrinos in the initial and final states in the flavomon emission/absorption processes have almost the same momentum $\mathbf{p}$. The energy and momentum conservation of the flavomon emission/absorption can be expressed as follows:%
\begin{align}
    \Omega - \omega - \mathbf{K}\cdot\mathbf{v} + \mu G^\alpha v_\alpha = 0,
    \label{eq:conservation-unshifted}
\end{align}
where  
\begin{align}
    G^\alpha = \int v^\alpha g(\omega, \mathbf{v})\,\frac{\mathrm{d}\omega\mathrm{d}\mathbf{v}}{4\pi}
    \label{eq:DLN-current}
\end{align}
is the DLN four-current.

In this work we revisit the Landau damping theory and explicitly demonstrate that the stable collective modes in the fast limit are indeed damped in realistic astrophysical environments. They are metastable at small wave numbers but become resonantly damped at large wave numbers, which is why, for example, the dispersion relation (DR) branch of the transverse mode exists only within a limited range of wave numbers in the fast limit. We also show that the elusive ``gapless'' modes that seemed to exist only within a short range of wave numbers \cite{Fiorillo:2025zio} can be harbingers of infinitely many collective modes with DR branches extending across the whole range of wave numbers. Although most of these DR branches are damped, the least damped resonant branches and the metastable branches can become unstable in certain ranges of wave numbers through inverse Landau damping.

The rest of the paper is organized as follows. In Sec.~\ref{sec:theory} we rederive Landau damping of the flavor excitations using the initial-value approach and demonstrate it with an explicit numerical example. In Sec.~\ref{sec:resonant} we introduce the concept of convolved DLN distributions and show the main features of the DR branches of the flavor excitations in the absence of flavor instabilities. In Sec.~\ref{sec:inverse-damping} we investigate the origin of the slow and fast flavor instabilities and show how they can be understood through the properties of the convolved DLN distributions. Finally, we summarize our results and conclude in Sec.~\ref{sec:summary}.

\section{Theory of Landau Damping}
\label{sec:theory}

\subsection{Equation of motion}

In the neutrino flavor isospin convention \cite{Duan:2005cp}, we define the effective neutrino oscillation frequency as
\begin{align}
    \omega = \left(\frac{\delta m^2}{2\varepsilon}\right)\cos (2\theta_\mathrm{v}),
\end{align}
where $\delta m^2$ is the mass-squared difference between the two neutrino mass eigenstates, $\varepsilon = |\mathbf{p}|$  and $-|\mathbf{p}|$ for the neutrino and the antineutrino, respectively, and $\theta_\mathrm{v}$ is the effective mixing angle in the two-flavor scenario. The DLN distribution \cite{Dasgupta:2009mg} is defined as
\begin{align}
    g(\omega, \mathbf{v}) \propto \frac{ \varepsilon^3}{|\omega|} \times
    \begin{cases}
        f_{\mathbf{p}}^{\nu_e} - f_{\mathbf{p}}^{\nu_\mu} & \text{if } \varepsilon>0, \\
        f_{\mathbf{p}}^{\bar\nu_e} - f_{\mathbf{p}}^{\bar\nu_\mu} & \text{if } \varepsilon<0,
    \end{cases}
\end{align}
with the normalization condition
\begin{align}
    G_0 = \int g(\omega, \mathbf{v})\,\frac{\mathrm{d}\omega\mathrm{d}\mathbf{v}}{4\pi} = 1.
    \label{eq:normalization}
\end{align}
We also define the flavor coherence distribution in terms of the oscillation frequencies as
\begin{align}
    \psi_{\omega, \mathbf{v}}(t, \mathbf{r}) \propto \frac{ \varepsilon^3}{|\omega|} \times
    \begin{cases}
        \psi_{\mathbf{p}} & \text{if } \varepsilon>0, \\
        \bar\psi_{\mathbf{p}} & \text{if } \varepsilon<0.
    \end{cases}
\end{align}

We focus on the regime where $\psi_{\omega, \mathbf{v}}$ is small and where $g(\omega, \mathbf{v})$ and the matter density are both constant in time and homogeneous in space. In this regime, the flavor evolution of a collisionless neutrino medium is governed by the linearized QKE (see, e.g., \cite{Izaguirre:2016gsx}):
\begin{align}
    v^\alpha\partial_\alpha \psi
    = -\mathrm{i} \omega \psi
    + \mathrm{i}\mu \!\int
    v^\beta v'_\beta
    (g' \psi - g \psi')\,
    \frac{\mathrm{d}\omega'\mathrm{d}\mathbf{v}'}{4\pi},
    \label{eq:eom}
\end{align}
where
\begin{subequations}
\label{eq:mu}
\begin{align}
    \mu &=
    \sqrt{2} G_\mathrm{F} (n_{\nu_e} - n_{\nu_\mu} - n_{\bar\nu_e} + n_{\bar\nu_\mu}) \\
    &= \sqrt{2} G_\mathrm{F}  \int (f_{\mathbf{p}}^{\nu_e} - f_{\mathbf{p}}^{\nu_\mu} - f_{\mathbf{p}}^{\bar\nu_e} + f_{\mathbf{p}}^{\bar\nu_\mu})\,\frac{\mathrm{d}^3p}{(2\pi)^3}
\end{align}
\end{subequations}
is the strength of the neutrino self-interaction potential. Here we have removed the matter effect by a co-rotating frame \cite{Duan:2005cp}. We also ignore a perturbation of order $\delta m^2\sin(2\theta_\mathrm{v})/\varepsilon$ that oscillates rapidly in the co-rotating frame. We suppress the dependence of $\psi$ and $g$ on $\omega$ and $\mathbf{v}$ in Eq.~\eqref{eq:eom} for brevity, and use the prime to denote the quantities that depend on the integration variables $\omega'$ and $\mathbf{v}'$.

The DLN distribution $g(\omega, \mathbf{v})$ and the self-interaction strength $\mu$ are not uniquely defined because only the product $\mu g(\omega, \mathbf{v})$ appears in Eq.~\eqref{eq:eom}. Later in the paper we will use DLN distributions that are normalized to $G_0\neq 1$. In this case, $\mu$ differs from the definition in Eq.~\eqref{eq:mu} by a factor of $G_0^{-1}$. In any case, we will measure all energies and frequencies in units of $\mu$ and set $\mu=1$ in the rest of the paper. 

\subsection{Laplace transform}

Following Landau \cite{Landau:1946jzi}, we consider a Fourier mode of (real) wave vector $\mathbf{K}$:
\begin{align}
    \psi_{\omega, \mathbf{v}}(t, \mathbf{r}) = \psi_{\omega, \mathbf{v}}(t)\,e^{\mathrm{i} \mathbf{K} \cdot \mathbf{r}}.
\end{align}
Its Laplace transform,
\begin{align}
    \tilde\psi_{\omega, \mathbf{v}}(s) = \int_0^\infty \psi_{\omega, \mathbf{v}}(t)\,e^{-st}\,\mathrm{d}t,
\end{align}
is defined in the half complex plane of $s$ to the right of all its singularities.

Multiplying both sides of Eq.~\eqref{eq:eom} by $e^{-st}$ and integrating over $t$ from $0$ to $\infty$, one obtains
\begin{align}
    \tilde\psi_{\omega, \mathbf{v}}(s)
= \frac{\mathrm{i}\psi_{\omega, \mathbf{v}}(0)
+ g(\omega, \mathbf{v}) v^\beta \tilde\Psi_\beta(s)}
{\mathrm{i} s - \omega - \mathbf{K} \cdot \mathbf{v} + v^\gamma G_\gamma},
\label{eq:phi-Phi}
\end{align}
where  
\begin{align}
    \tilde\Psi^\alpha(s) = \int v^\alpha \tilde\psi_{\omega, \mathbf{v}}(s) \,\frac{\mathrm{d}\omega\mathrm{d}\mathbf{v}}{4\pi}
\end{align}
is the Laplace transform of
\begin{align}
    \Psi^\alpha(t) = \int v^\alpha \psi_{\omega, \mathbf{v}}(t) \,\frac{\mathrm{d}\omega\mathrm{d}\mathbf{v}}{4\pi}.
\end{align}

Multiplying both sides of Eq.~\eqref{eq:phi-Phi} by $v^\alpha$ and integrating over the phase space, one obtains the self-consistency equation
\begin{align}
   [\delta^\alpha_\beta - \chi^{\alpha}_{\beta}(s)] \tilde\Psi^\beta(s) = A^\alpha(s),
    \label{eq:ssc}
\end{align}
where
\begin{align}
    \chi^{\alpha}_{\beta}(s) = \int \frac{v^\alpha v_\beta g(\omega, \mathbf{v})}{\mathrm{i} s - \omega - \mathbf{K}\cdot\mathbf{v} + v^\gamma G_\gamma} \,\frac{\mathrm{d}\omega\mathrm{d}\mathbf{v}}{4\pi}
    \label{eq:chi}\\
    \intertext{and}
    A^\alpha(s) = \int \frac{\mathrm{i} v^\alpha \psi_{\omega, \mathbf{v}}(0)}{\mathrm{i} s - \omega - \mathbf{K}\cdot\mathbf{v} + v^\gamma G_\gamma} \,\frac{\mathrm{d}\omega\mathrm{d}\mathbf{v}}{4\pi}.
    \label{eq:A}
\end{align}

Although $g(\omega, \mathbf{v})$ and $\psi_{\omega, \mathbf{v}}(0)$ are defined on the real axis of $\omega$, we assume that they can be analytically continued to the whole complex plane and thus are entire functions of $\omega$. We also assume that they vanish sufficiently fast as $\omega\to\pm\infty$. Equations~\eqref{eq:chi} and \eqref{eq:A} define $\chi^\alpha_\beta(s)$ and $A^\alpha(s)$ in the right half of the complex plane of $s$. They can be analytically continued to the left half of the complex plane as follows.  

\begin{figure}[htb]
    \begin{tabular}{c@{\hspace{12pt}}c}
        \includegraphics[scale=0.45]{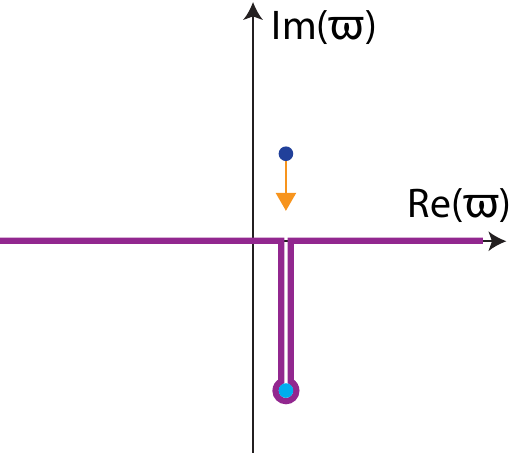} &
        \includegraphics[scale=0.45]{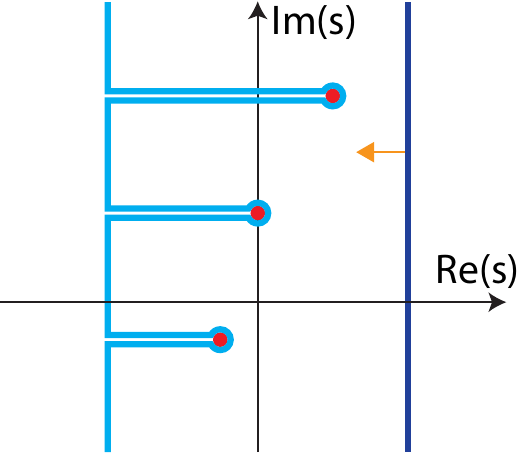}
    \end{tabular}
    \caption{Left: the contour of the $\varpi$-integration in Eq.~\eqref{eq:Z2} is deformed as the pole (shown as the solid dot) of the integrand descends across the real axis.
    Right: the contour of the $s$-integration for the inverse Laplace transform can be deformed from the right half plane to the left half plane except around the poles of $\tilde\Psi^\alpha(s)$ (shown as the solid dots).}
    \label{fig:contours}
\end{figure}

Both Eqs.~\eqref{eq:chi} and \eqref{eq:A} involve integrals that can be recast in the form of the generalized plasma dispersion function:
\begin{subequations}
        \label{eq:Z}
    \begin{align}
    Z_H(s) &= \int   \frac{h(\omega, \mathbf{v})}{\mathrm{i} s - \omega - \mathbf{K}\cdot\mathbf{v} + v^\alpha G_\alpha}
    \,\frac{\mathrm{d}\omega\mathrm{d}\mathbf{v}}{4\pi}
    \label{eq:Za}\\
    &= \int_{-\infty}^{\infty}  \frac{H(\varpi)\, \mathrm{d}\varpi}{\mathrm{i} s + G_0 - \varpi},
\end{align}
\end{subequations}
where
\begin{align}
    H(\varpi) = \int h(\varpi - (\mathbf{K} + \mathbf{G})\cdot\mathbf{v}, \mathbf{v})\,\frac{\mathrm{d}\mathbf{v}}{4\pi}
\end{align}
is an entire function of $\varpi$.
Equation~\eqref{eq:Z} defines $Z_H(s)$ for $\mathrm{Re}(s) > 0$. 
To analytically continue $Z_H(s)$ to the left half of the complex plane, Landau deformed the integration contour to pass below the simple pole of the integrand (see the left panel of Fig.~\ref{fig:contours}) so that
\begin{subequations}
    \label{eq:Z2}
\begin{align}
    Z_H(s) &= \int_\mathcal{L}  \frac{H(\varpi)\, \mathrm{d}\varpi}{\mathrm{i} s + G_0 - \varpi}\\
    &=
    \int_{-\infty}^{\infty}  \frac{H(\varpi)\, \mathrm{d}\varpi}{\mathrm{i} s + G_0 - \varpi}
     - 2\pi \mathrm{i} \Theta(-\mathrm{Re}(s)) H(\mathrm{i} s + G_0), 
\end{align}
\end{subequations}
is an entire function of $s$, where $\mathcal{L}$ represents the Landau contour, and $\Theta(x)$ is the Heaviside step function.

With the definition of $Z_H(s)$ given by Eq.~\eqref{eq:Z2}, $\tilde{\Psi}^\alpha(s)$ is defined on the whole complex plane through Eq.~\eqref{eq:ssc} except at its poles
\begin{align}
    s_k = - \mathrm{i}\Omega_k + \Gamma_k,
\end{align}
which are the roots of
\begin{align}
    \det[\delta^\alpha_\beta - \chi^{\alpha}_{\beta}(s_k)] = 0.
    \label{eq:DR}
\end{align}

\subsection{Asymptotic behavior}
\label{sec:asymptotic}

For $t>0$, one can perform the inverse Laplace transform to obtain
\begin{align}
    \Psi^\alpha(t)
    = \frac{1}{2\pi\mathrm{i}} \int_{a - \mathrm{i}\infty}^{a + \mathrm{i}\infty} \tilde\Psi^\alpha(s)\,e^{st}\,\mathrm{d}s,
    \label{eq:Phi}
\end{align}
where $a$ is larger than all $\Gamma_k$. The contour of the $s$-integration in Eq.~\eqref{eq:Phi} can be deformed to the left half of the complex plane except around $s_k$ (the right panel of Fig.~\ref{fig:contours}). This new integration contour shows that the asymptotic behavior of $\Psi^\alpha(t)$ at large $t$ is given by the contribution from the pole of $\tilde\Psi^\alpha(s)$ that has the largest real part, $s_\star = \Gamma_\star - \mathrm{i} \Omega_\star$, and
\begin{align}
    \Psi^\alpha(t) \sim e^{\Gamma_\star t - \mathrm{i} \Omega_\star t} \quad \text{as } t\to\infty.
\end{align}

Suppose $\tilde\Psi^\alpha(s)$ has only one pole $s_\star$. If it is in the right half of the complex plane, both $\Psi^\alpha(t)$ and $\psi_{\omega, \mathbf{v}}(t)$ grow exponentially with rate $\Gamma_\star$ and oscillate with frequency $\Omega_\star$. This pole corresponds to an unstable collective mode of the neutrino medium. If the pole is in the left half of the complex plane, however, $\psi_{\omega, \mathbf{v}}(t)$ at large $t$ is dominated by the pole of the denominator in Eq.~\eqref{eq:phi-Phi} at $s = -\mathrm{i} \Omega$ with $\Omega = \omega + \mathbf{K}\cdot\mathbf{v} - v^\alpha G_\alpha$, and
\begin{align}
    \psi_{\omega, \mathbf{v}}(t) \sim e^{-\mathrm{i} \Omega t} \quad \text{as } t\to\infty.
\end{align}
This is a non-collective mode of the neutrino medium \cite{Capozzi:2019lso} which corresponds to a Case--van Kampen mode in a plasma \cite{VanKampen:1955wh,Case:1959}. Nevertheless, $\Psi^\alpha(t)$ decays with rate $|\Gamma_\star|$ at large $t$. This phenomenon is known as Landau damping \cite{Landau:1946jzi} and is a result of the phase mixing of the non-collective modes \cite{Fiorillo:2024bzm}. 

The theory of Landau damping was first applied to the DRs of pure fast modes in Ref.~\cite{Fiorillo:2024bzm} and extended to neutrino gases with continuous frequency distributions in Ref.~\cite{Fiorillo:2025zio}. In contrast to our approach, Ref.~\cite{Fiorillo:2025zio} performs the integration in Eq.~\eqref{eq:Za} over the neutrino velocity $\mathbf{v}$ first and without a change of variables, leading to a result with branch points in the vacuum frequency $\omega$ that arise from the finite range of the integration over $\mathbf{v}$. The subsequent integration over $\omega$ is performed along a contour that, when $\mathrm{Re}(s) < 0$, crosses the corresponding branch cuts, rather than going around them as it should. Although the two approaches give identical results for unstable modes, the approach in Ref.~\cite{Fiorillo:2025zio} can give incorrect results for damped modes. See Fig.~\ref{fig:triple-gaussian} and the discussion at the end of Sec.~\ref{sec:sfi} for an example that illustrates the difference between the two approaches.

\subsection{Landau damping}
\label{sec:damping}

To illustrate Landau damping, we consider a homogeneous and isotropic neutrino gas with the DLN distribution
\begin{align}
    g(\omega, \mathbf{v}) = g(\omega).
\end{align}
For now, we restrict our attention to homogeneous modes, which have $\mathbf{K} = \mathbf{0}$. In this case, the DLN four-current is $G^\alpha = (1,\mathbf{0})$, and Eq.~\eqref{eq:DR} reduces to
\begin{align}
    (1 - \chi^0_0)(1 - \chi^1_1)(1 - \chi^2_2)(1 - \chi^3_3) = 0
\end{align}
or
\begin{align}
    \mathcal{D}(\Omega+\mathrm{i}\Gamma) 
    \equiv \int_{\mathcal{L}} \frac{g(\omega)\,\mathrm{d}\omega}{\Omega + \mathrm{i}\Gamma - \omega}
    - b = 0,
    \label{eq:DR-homo}
\end{align}
where $b=1$ for the monopole mode and $-3$ for the three dipole modes. The monopole mode $(\Psi^0, \mathbf{0})$ preserves the isotropy of the gas, and the dipole modes $(0, \bm{\Psi})$ break the isotropy \cite{Duan:2013kba}.
As usual, we have redefined the frequency and wave vector of a collective mode as
\begin{align}
    \Omega + \mu G_0 \to \Omega
    \qquad\text{and}\qquad
    \mathbf{K} + \mu \mathbf{G} \to \mathbf{K}
\end{align}
with which Eq.~\eqref{eq:conservation-unshifted} becomes
\begin{align}
    \Omega - \mathbf{K} \cdot \mathbf{v}  = \omega.
    \label{eq:conservation}
\end{align}
Contrary to the convention adopted by most of the literature, we always use capital letters to denote the frequencies and wave vectors of the collective modes, and we reserve the lowercase letters for those of the neutrinos.

Equation~\eqref{eq:DR-homo} shows that the DR of the dipole modes can be obtained from the same equation as for the monopole mode by multiplying the DLN distribution by a factor of $-1/3$. This factor stems from the current-current structure of the weak interaction \cite{Duan:2013kba}. It can also be partly understood in the flavomon picture as follows. In the flavomon emission/absorption processes in Eq.~\eqref{eq:processes}, both the incoming and outgoing neutrinos are in the same spin state $|s=1/2, m_s=-1/2; \mathbf{v}\rangle$, where $\mathbf{v}$ indicates the quantization axis of the spin state. The monopole flavomon represented by $(\Psi^0, \mathbf{0})$ has spin $s=0$. Therefore, the conservation of angular momentum along the direction of $\mathbf{v}$ does not add any further constraint. The dipole flavomon represented by $(0,0,0,\Psi^3)$, however, is in the spin state $|s=1, m_s=0; \hat{\mathbf{z}}\rangle$. The conservation of angular momentum along $\mathbf{v}$ implies that the emission/absorption rate of this flavomon is proportional to 
\begin{align}
    |\langle  s=1, m_s=0;  \hat{\mathbf{z}} |  s=1, m_s=0; \mathbf{v}\rangle |^2 = u^2,
    \label{eq:dipole-factor}
\end{align}
where 
\begin{align}
    u = \mathbf{v}\cdot\hat{\mathbf{z}}.
\end{align}
The factor in Eq.~\eqref{eq:dipole-factor}, after averaging over the neutrino velocity $\mathbf{v}$, gives $1/3$ for the dipole modes. Since the medium is isotropic, each dipole flavomon is in resonance with $1/3$ of the neutrinos that are in resonance with the monopole flavomon. 

It is known that there is no flavor instability if $g(\omega)$ has no zero crossings. This can be seen directly from Eq.~\eqref{eq:DR-homo} as follows. A flavor instability requires $\Gamma > 0$ and $\mathrm{Im}(\mathcal{D}) = 0$. However, a direct calculation shows that
\begin{align}
    \mathrm{Im}(\mathcal{D}) = -\Gamma \int_{-\infty}^{\infty} \frac{g(\omega)\,\mathrm{d}\omega}{(\Omega - \omega)^2 + \Gamma^2} 
    \quad\text{for } \Gamma > 0,
\end{align}
which cannot be zero if $g(\omega)$ has no zero crossing.
In addition, because 
\begin{align}
    \mathcal{D}_\mathrm{I}(\Omega) \equiv \mathrm{Im}[\mathcal{D}(\Omega)] = -\pi g(\Omega),
\end{align}
the frequency $\Omega$ of a stable mode (with $\Gamma=0$) must be a zero point of the DLN distribution. 

For a tiny but nonzero $\Gamma$, one can write
\begin{align}
   \mathcal{D}(\Omega+\mathrm{i}\Gamma) 
   \approx   \mathcal{D}_\mathrm{R}(\Omega)
   + \mathrm{i}\Gamma \partial_{\Omega} \mathcal{D}_\mathrm{R}(\Omega)
   + \mathrm{i} \mathcal{D}_\mathrm{I}(\Omega),
\end{align}
where
\begin{align}
    \mathcal{D}_\mathrm{R}(\Omega) = \mathrm{Re}[\mathcal{D}(\Omega)] 
    = \mathrm{PV}\int_{-\infty}^{\infty} 
    \frac{g(\omega)\,\mathrm{d}\omega}{\Omega - \omega}
    - b
\end{align}
with ``PV'' denoting the Cauchy principal value of the integral. Therefore, the frequency and growth rate of a nearly stable mode can be obtained from
\begin{align}
    \mathcal{D}_\mathrm{R}(\Omega) \approx 0 
    \quad\text{and}\quad
    \Gamma \approx -\frac{\mathcal{D}_\mathrm{I}(\Omega)}{\partial_{\Omega} \mathcal{D}_\mathrm{R}(\Omega)}
    \label{eq:Gamma-small}
\end{align}
(unless $\partial_{\Omega} \mathcal{D}_\mathrm{R}(\Omega) = 0$). 
Further assuming $|\Omega| \sim \mu$, which is much larger than the typical vacuum oscillation frequency $\omega$ of the neutrinos, one has
\begin{subequations}
\begin{align}
    \Omega_\mathrm{m} \approx G_0 = 1 &\quad \text{and} \quad \Gamma_\mathrm{m} \approx  \frac{ -\pi g(\Omega_\mathrm{m})}{G_0 / \Omega_\mathrm{m}^2}
    \approx -\pi g(1)
    \label{eq:monopole}\\
    \intertext{for the monopole mode, and}
    \Omega_\mathrm{d} \approx - \frac{G_0}{3} = -\frac{1}{3} 
    &\quad \text{and} \quad 
    \Gamma_\mathrm{d} \approx  \frac{ -\pi g(\Omega_\mathrm{d})}{G_0 / \Omega_\mathrm{d}^2}
    \approx -\frac{\pi }{9} g\left(-\frac{1}{3}\right)
    \label{eq:dipole}
\end{align}
\end{subequations}
for the dipole modes.

As a concrete example, we consider a $\nu_e$ gas with the Fermi--Dirac distribution%
\footnote{When solving Eq.~\eqref{eq:DR-homo}, we replace $\Theta(x)$ in Eq.~\eqref{eq:FD} with $\Theta(\mathrm{Re}(x))$. Although this makes the function non-entire, we find numerically that it does not significantly affect the calculations of the Landau damping rates.}
\begin{align}
    g(\omega)
    =  \frac{2|\beta|^3}{3\zeta(3)} 
    \frac{\omega^{-4}}{e^{\beta/\omega} + 1}
    \Theta\left(\frac{\omega}{\delta m^2}\right),
    \label{eq:FD}
\end{align}
where $\beta = \delta m^2 \cos(2\theta_\mathrm{v})/2 T$ with $T$ being the temperature parameter of the neutrino gas. Note that $g(\omega)$ is nonzero only at $\omega>0$ ($\omega<0$) for the normal (inverted) mass ordering with $\delta m^2 >0$ and $\beta >0$ ($\delta m^2 < 0$ and $\beta < 0$). 

For $T \sim 1\,\mathrm{MeV}$, $\delta m^2 = \delta m^2_\mathrm{atm}$, and $\mu\sim 10^{5}\,\mathrm{km}^{-1}$, one has $\beta/\mu \sim 6\times 10^{-5}$ for which Eq.~\eqref{eq:monopole} predicts $\Gamma_\mathrm{m} \sim -2.2\times 10^{-13}$, which is too small to be observed in numerical simulations. For illustration purposes, we will use $\beta/\mu = \pm 0.1$ in the following numerical examples. 

For $\beta = 0.1$, Eq.~\eqref{eq:DR-homo} has roots $(\Omega_\mathrm{m}, \Gamma_\mathrm{m}) \approx (1.05, -6.88\times10^{-4})$ for the monopole mode and $(\Omega_\mathrm{d}, \Gamma_\mathrm{d}) \approx (-0.29, 0)$ for the dipole modes. For $\beta = -0.1$, the roots are $(\Omega_\mathrm{m}, \Gamma_\mathrm{m}) \approx (0.96, 0)$ and $(\Omega_\mathrm{d}, \Gamma_\mathrm{d}) \approx (-0.38, -3.88\times10^{-3})$, respectively. For comparison, Eqs.~\eqref{eq:monopole} and \eqref{eq:dipole} give approximate values of $\Gamma_\mathrm{m} \approx -8.28\times10^{-4}$ for $\beta=0.1$ and $\Gamma_\mathrm{d} \approx -6.67\times10^{-3}$ for $\beta=-0.1$.

\begin{figure}[tb]
    \includegraphics[width=\linewidth, trim=8 10 5 5, clip=true]{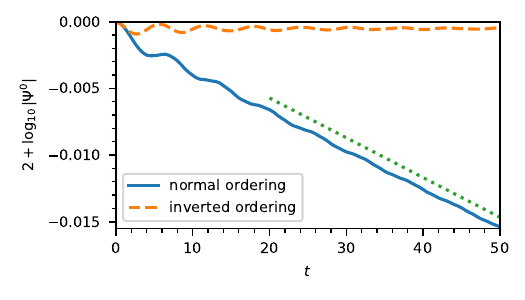}
    \caption{The monopole moment of the neutrino coherence $|\Psi^0(t)|$ in a homogeneous and isotropic $\nu_e$ gas with a small initial flavor coherence. The solid and dashed lines represent the numerical results for the normal and inverted neutrino mass orderings, respectively. The dotted line represents an exponential decay with the rate predicted by Landau damping for the normal mass ordering.}
    \label{fig:example}
\end{figure}

The \texttt{f2i} model of the \texttt{NuGas} Python package \cite{nugas} is capable of tracking the monopole mode in a homogeneous and isotropic neutrino gas. Using this software, we performed numerical simulations for a $\nu_e$ gas with the Fermi--Dirac distribution $g(\omega)$ with $\beta=\pm 0.1$ and with a small initial flavor coherence $\psi_{\omega,\mathbf{v}}(0) = 10^{-2}\, g(\omega)$. The results are shown in Fig.~\ref{fig:example}. Indeed, the amplitude of the monopole mode stays approximately constant for the inverted mass ordering and decays exponentially with the rate predicted by Landau damping for the normal ordering.

\section{Resonant Collective Modes}
\label{sec:resonant}

In illustrating Landau damping in Sec.~\ref{sec:damping}, we have assumed $|\Omega|$ to be much larger than the typical oscillation frequency $\omega$ of the neutrinos. These modes are stable in the fast limit but become metastable when $g(\Omega)$ is small but nonzero. It turns out that there can also exist infinitely many other collective modes that are more strongly damped than these metastable modes. To demonstrate their existence, we consider a homogeneous and isotropic neutrino gas with a Gaussian DLN distribution:
\begin{align}
    g(\omega, \mathbf{v}) =  g(\omega) = \phi(\omega;0,\sigma),
    \label{eq:single-gaussian}
\end{align}
where
\begin{align}
    \phi(\omega; \omega_0, \sigma) = \frac{1}{\sqrt{2\pi}\sigma} e^{-(\omega - \omega_0)^2/2\sigma^2},
\end{align}
and $0<\sigma\ll 1$ is the width of the frequency DLN distribution $g(\omega)$. 
We choose the Gaussian distribution because it is an entire function and the associated results can be understood analytically. We do not expect our general conclusions to depend on this specific choice.
The distribution is centered at $\omega_0 = 0$ for convenience. One can always shift the center of $g(\omega)$ by changing to an appropriate co-rotating frame, which redefines $\Omega$.

\subsection{Homogeneous modes}
\label{sec:homo}

For $\mathbf{K}=\mathbf{0}$, the DR equation~\eqref{eq:DR-homo} can be recast in the form
\begin{align}
    w\left(\frac{1 + \mathrm{i} s}{\sqrt{2}\sigma}\right)
    = \mathrm{i} \sqrt{\frac{2}{\pi}} \sigma b,
    \label{eq:DR-homo2}
\end{align}
where
\begin{align}
    w(z) &= e^{-z^2} \mathrm{erfc}(-\mathrm{i} z) 
    = \frac{\mathrm{i}}{\pi} \int_\mathcal{L} \frac{e^{-t^2}\,\mathrm{d}t}{z - t} 
\end{align}
is the Faddeeva function \cite{faddeyeva_terentev_1961}.%
\footnote{A related function is the plasma dispersion function $Z(\xi)=\mathrm{i}\sqrt\pi w(\xi)$, which is widely used in plasma physics \cite{fried_conte_1961}.}
In the limit $|\sigma b| \ll1$, two ladders of solutions to Eq.~\eqref{eq:DR-homo2} are approximately given by the Faddeeva zeros: 
\begin{align}
    1 + \mathrm{i} s_k = \Omega_k + \mathrm{i}\Gamma_k
    \approx \sqrt{2}\sigma \zeta_k
    \quad(k = \pm 1, \pm 2, \ldots),
\end{align}
where $w(\zeta_k) = 0$. (The monopole and dipole solutions differ slightly for small but finite $\sigma$.)
Faddeeva zeros come in pairs with opposite real parts and equal imaginary parts. The first three pairs of the Faddeeva zeros are $\zeta_{\pm1} \approx \pm1.99 - 1.35 \mathrm{i}$, $\zeta_{\pm2} \approx \pm2.70 - 2.18 \mathrm{i}$, and $\zeta_{\pm3} \approx \pm 3.24 - 2.78 \mathrm{i}$.

\begin{figure*}[htbp]
\includegraphics[width=\textwidth, trim=0 0 0 0, clip]{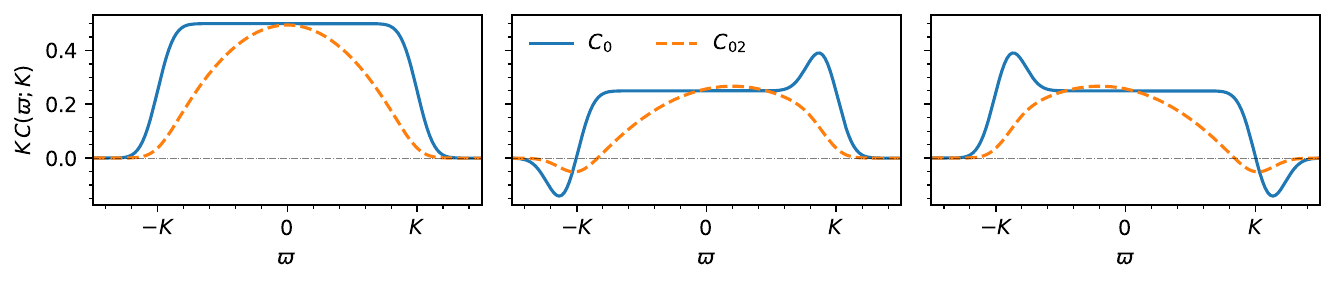}
\caption{Examples of the convolved DLN distributions $C_0(\varpi; K)$ (solid curves) and $C_{02}(\varpi; K)$ (dashed curves) for three different frequency DLN distributions $g(\omega)$. Left: $g(\omega)$ is a positive, normalized Gaussian distribution centered at $\omega = 0$; Middle: a smaller negative Gaussian is added to the left of the positive one; Right: the negative Gaussian is placed to the right of the positive one.}
\label{fig:conv}
\end{figure*}

Compared to the \emph{metastable modes}, these modes are more strongly damped with $\Gamma_k \sim -\sqrt{(2|k| - 1/4)\pi}\sigma$ \cite{NIST:DLMF}. Their damping rates are not determined by $g(\Omega_k)$ as in Eq.~\eqref{eq:Gamma-small} because the DLN distribution changes significantly over the resonance frequency range $(\Omega_k - |\Gamma_k|, \Omega_k + |\Gamma_k|)$ of these modes.
These \emph{resonant modes} correspond to the ``gapless modes'' in Ref.~\cite{Fiorillo:2025zio} because $\Omega_k \sim \pm \sqrt{(2|k| - 1/4)\pi}\sigma$ vanish in the fast limit $\sigma\to0$, while the metastable modes correspond to the ``gapped modes'' that have $\Omega$ of order $\mu\gg\sigma$ and survive in the fast limit. However, for a finite spread $\sigma$, there is an infinite number of gapless modes that can have $|\Omega_k|$ even larger than those of the gapped modes. Nevertheless, the resonant modes are still gapless in the sense that there is a significant overlap between their resonance frequency range and the ``effective support'' of the DLN distribution, i.e., the frequency range where $g(\omega)$ is appreciable. In contrast, $\Omega$ of a gapped/metastable mode is separated from the effective support of the DLN distribution.

\subsection{Convolved DLN distributions}

The inhomogeneous modes with $\mathbf{K} = K \hat{\mathbf{z}}$ ($K>0$) fall into two categories according to their symmetry properties: the longitudinal modes that are axially symmetric about the $z$ axis and the transverse modes that break the axial symmetry \cite{Izaguirre:2016gsx}. 

Much of the insight we have gained about the homogeneous metastable and resonant modes still applies, except that their properties are determined by the convolved DLN distributions 
\begin{align}
    C_n(\varpi; K) = \int_{-1}^{1} u^n g(\varpi - K u)\,\frac{\mathrm{d}u}{2}
\end{align}
instead of the frequency DLN distribution $g(\omega)$ itself.
In particular, $C_0(\varpi; K)$ is proportional to the DLN of the neutrinos that are in resonance with the collective mode of frequency $\Omega=\varpi$ and wave number $K$ when $\Gamma$ is small. Therefore, if there is no other constraint, a necessary condition for the existence of a flavor instability at wave number $K$ is that $C_0(\varpi; K)$ has a zero crossing in terms of $\varpi$. One expects $\Gamma\propto -C_0(\Omega; K)$ if a collective mode has a small damping or growth rate and if there is no constraint from the conservation of angular momentum. As we will see shortly, 
\begin{align}
    C_{02}(\varpi; K) & = \int_{-1}^{1} (1-u^2) g(\varpi - K u)\,\frac{\mathrm{d}u}{2}
\end{align}
is the convolved DLN distribution that actually determines the properties of the transverse modes.

In the left panel of Fig.~\ref{fig:conv}, we show $C_0(\varpi; K)$ with a Gaussian frequency DLN distribution given by Eq.~\eqref{eq:single-gaussian}. It is essentially a box distribution of width $2K$ and height $1/2K$ with its edges smoothed by the convolution over a range of width $\sim 2\sigma$. The edges of the distribution at $\varpi=\pm K$ represent the boundaries of the ``light cone''. If the properties of the collective modes are determined by $C_0(\varpi; K)$ alone, then those with $|\Omega| - K \gg |\Gamma| + \sigma$ are not resonant with a significant number of neutrinos and, therefore, must be metastable, while those with $|\Omega| \lesssim K + |\Gamma| + \sigma$ are resonant modes. In the same figure we also show $C_{02}(\varpi; K)$, which is an inverted parabola of width $2K$ and height $1/2K$ and with its edges smoothed by the convolution.

\subsection{Transverse modes}
\label{sec:transverse}

In the flavomon picture, a transverse flavomon is in the spin quantum state $|s=1, m_s=\pm 1; \hat{\mathbf{z}}\rangle$. The conservation of angular momentum implies that its emission/absorption rate is proportional to 
\begin{align}
    \left|\langle s=1, m_s=0; \mathbf{v} |s =1, m_s=\pm 1; \hat{\mathbf{z}}\rangle \right|^2 = \frac{1 - u^2}{2},
    \label{eq:T-factor}
\end{align}
where $\mathbf{v}$ is the velocity of the neutrinos that emit or absorb the flavomon.
Therefore, one expects that the properties of the transverse modes are determined by $C_{02}(\varpi; K)$ instead of $C_0(\varpi; K)$. Indeed, the DR of the transverse modes is given by
\begin{align}
    \mathcal{D}_\mathrm{T}(\Omega + \mathrm{i}\Gamma) = Z_0 - Z_2 + 2 = 0,
    \label{eq:DR-T}
\end{align}
where
\begin{align}
    Z_n(\Omega + \mathrm{i}\Gamma) = \int_\mathcal{L} \frac{C_n(\varpi; K) \,\mathrm{d}\varpi}{\Omega + \mathrm{i} \Gamma - \varpi}.
\end{align}
Equation~\eqref{eq:DR-T} shows that the DR of the transverse modes can be obtained from the same equation as for the monopole mode [Eq.~\eqref{eq:DR-homo} with $b=1$] by replacing $g(\omega)$ with $(-1/2) C_{02}(\varpi; K)$.
The minus sign stems from the spatial part of the weak current coupling in $1-\mathbf{v}\cdot\mathbf{v}'$, and the factor $(1-u^2)/2$ in $(1/2) C_{02}(\varpi; K)$  is what we have shown in Eq.~\eqref{eq:T-factor}.

\begin{figure*}[htbp]
\includegraphics[width=\textwidth]{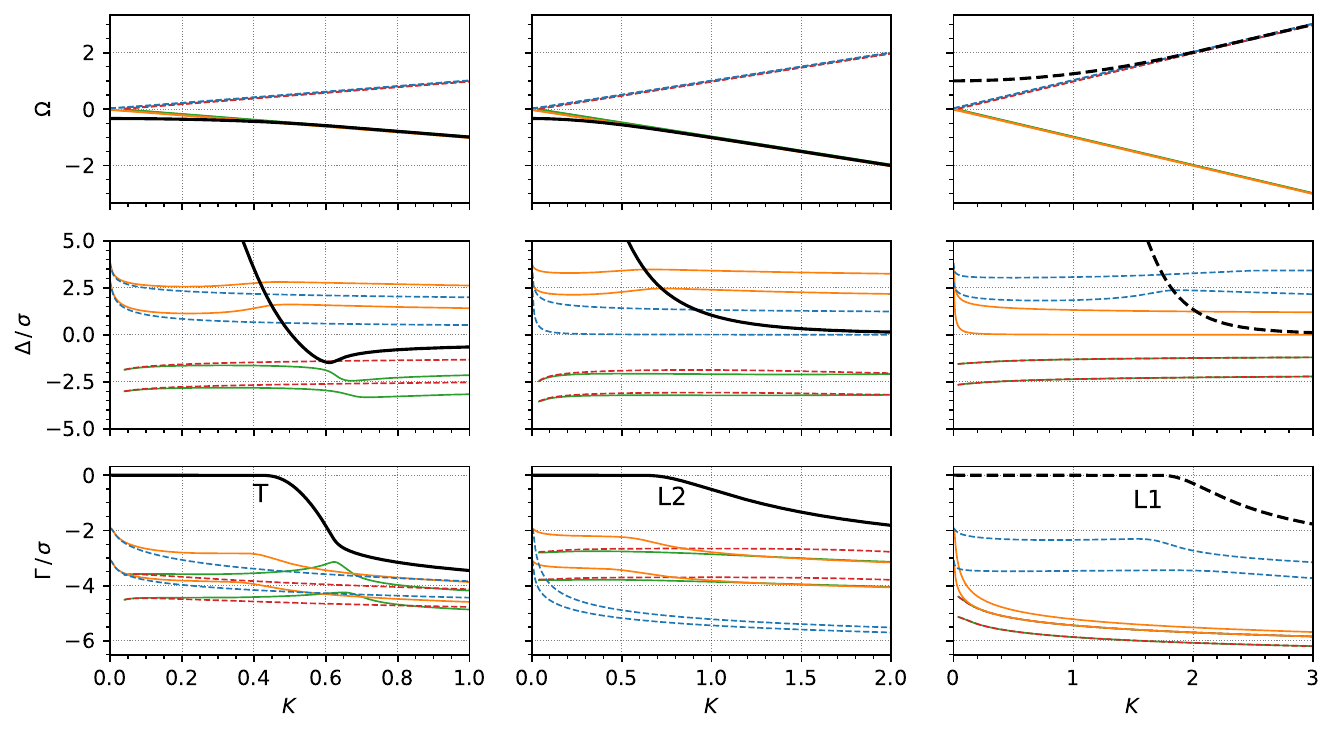}
\caption{The DRs of the transverse (left column) and longitudinal (middle and right columns) branches of the collective modes in an isotropic neutrino gas with the Gaussian DLN distribution $g(\omega) = \phi(\omega;0,\sigma)$ with $\sigma=0.01$. Top row: frequencies $\Omega$ of the collective modes as functions of the wave number $K$. Middle row: distance from the light cone $\Delta=|\Omega| - |K|$. Superluminal and subluminal modes have $\Delta > 0$ and $\Delta < 0$, respectively. Bottom row: growth rates $\Gamma$. The thick curves in the left, middle, and right columns represent the T, L2, and L1 branches that are metastable at small $K$, and the thin curves represent the branches of the collective modes that are always resonant. The dashed and solid curves represent the branches with positive and negative $\Omega$, respectively. Various colors represent different ladders of the resonant branches. Only the first two resonant branches are shown in each ladder. The two subluminal resonant branches in each pair have essentially the same values of $\Delta$ and $\Gamma$ and are difficult to distinguish in the middle and bottom panels of the right column.} 
\label{fig:single-gaussian}
\end{figure*}

Because of the similarity between Eqs.~\eqref{eq:DR-T} and \eqref{eq:DR-homo}, we expect $\Omega(K)$ and $\Gamma(K)$ of the transverse modes at $K>0$ to behave similarly to those of the dipole modes at $K=0$ (which also have a negative sign stemming from the coupling of the spatial part of the weak current). This is indeed the case, as shown by the example in the left column of Fig.~\ref{fig:single-gaussian}, which demonstrates some important features of the transverse modes:
\begin{itemize}
    \item Two gapped dipole modes at $K=0$ expand into two degenerate transverse branches (represented by the same thick solid curves) that remain metastable at $K\lesssim 1/2$. At larger $K$, their frequency $\Omega$ enters the effective support of $C_{02}(\varpi; K)$ when $|\Omega| - K \lesssim \sigma$. As a result, they are resonant with a significant number of neutrinos, and their damping rate increases rapidly with $K$. 
    We call these two degenerate branches the T branches, following the convention in Ref.~\cite{Fiorillo:2025kko}.
    \item As shown in Sec.~\ref{sec:homo}, there are two ladders of gapless dipole modes for each gapped dipole mode at $K=0$. These gapless modes expand into two ladders of branches that remain resonant for all $K$. They are ``superluminal'' in the sense that they are outside the light cone defined by $\Omega=\pm K$. In Fig.~\ref{fig:single-gaussian}, we have also plotted the distance of the collective modes from the light cone
    \begin{align}
        \Delta = |\Omega| - |K|
    \end{align}
    which is positive for superluminal branches.
    \item In addition to the superluminal resonant branches, there also exists a pair of ladders of ``subluminal'' resonant branches with $\Delta < 0$ for each T branch. At small $K$, the subluminal modes in each pair have similar damping rates and magnitudes of $\Omega$ but opposite signs of $\Omega$. They seem to collide near $K=0$ and become difficult to track at even smaller $K$.
    
\end{itemize}

\subsection{Longitudinal modes}
\label{sec:longitudinal}

Compared to the transverse modes, the longitudinal modes are more complex, and their DR is given by
\begin{align}
    \mathcal{D}_\mathrm{L} = (Z_0- 1)(Z_2 + 1) - Z_1^2 = 0 .
\end{align}
This is because a longitudinal flavomon is a superposition of two spin states $|s=0, m_s=0 ; \hat{\mathbf{z}} \rangle$ and $|s=1, m_s=0;  \hat{\mathbf{z}}\rangle$. For the near-luminal flavomons with $|\Omega|\approx K \gg\sigma$, however, the energy-momentum conservation given by Eq.~\eqref{eq:conservation} implies that they can be in resonance only with the neutrinos of $u\approx \pm 1$, i.e., $\mathbf{v}\approx\pm\hat{\mathbf{z}}$. In this case, the conservation of angular momentum does not add any further constraint, and the damping/growth rates of the near-luminal longitudinal modes should be largely determined by $C_0(\varpi; K)$. (See Appendix~\ref{sec:luminal} for more discussion.)

In the right two columns of Fig.~\ref{fig:single-gaussian} we show the DRs of the longitudinal modes for a Gaussian frequency DLN distribution. From this example one can see that the longitudinal modes have features similar to those of the transverse modes just discussed, except with twice as many branches:  
\begin{itemize}

    \item The gapped monopole mode and one of the gapped dipole modes at $K=0$ expand into two longitudinal branches that remain metastable at small $K$. Although they remain superluminal for all $K$, they become resonant when $|\Omega| - K \lesssim \sigma$ because their frequencies have entered the effective support of $C_0(\varpi; K)$. We call these two branches the L1 and L2 branches, respectively, following the convention in Ref.~\cite{Fiorillo:2025kko}.
    
    \item The ladders of gapless monopole and dipole modes at $K=0$ expand into ladders of superluminal branches that remain resonant for all $K$. We show the first two resonant branches in each ladder in the right or middle column of Fig.~\ref{fig:single-gaussian} depending on whether they reduce to the gapless monopole modes at $K=0$ as L1 does or to the gapless dipole modes as L2 does.
    
    \item As in the transverse case, there exist pairs of subluminal branches that remain resonant for all $K$ and collide near $K=0$. We also show the first two resonant branches in each of these ladders in the middle and right columns of Fig.~\ref{fig:single-gaussian}, although the classification of the subluminal branches is arbitrary because we do not track them all the way to $K=0$.
\end{itemize}

\section{Inverse Landau Damping}
\label{sec:inverse-damping}

\begin{figure*}[tbp]
\includegraphics[width=\textwidth]{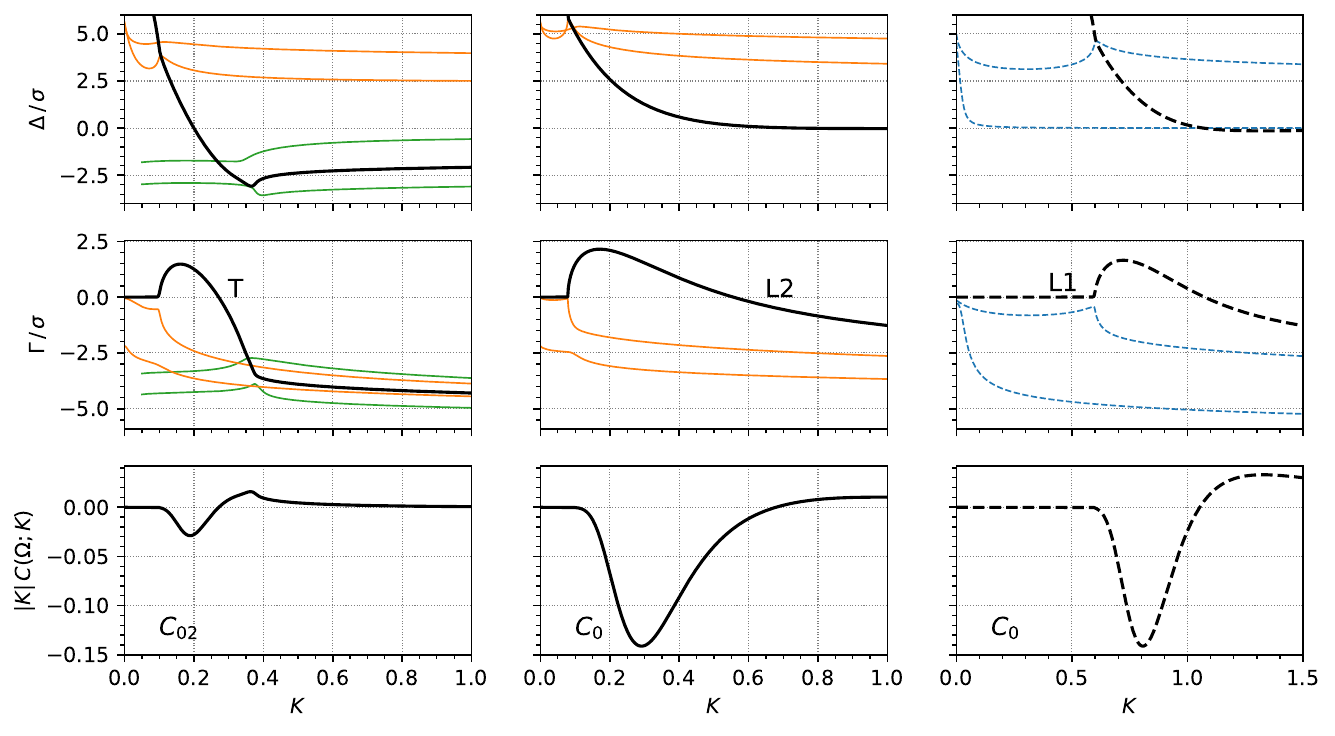}
\caption{Similar to Fig.~\ref{fig:single-gaussian} but for double-Gaussian DLN distributions $g(\omega) = \phi(\omega; 0, \sigma) - 0.5\,\phi(\omega; \delta\omega, \sigma)$ with $\sigma=0.01$, where $\delta\omega=-2\sigma$ for the left two columns (with the T and L2 branches) and $\delta\omega=+2\sigma$ for the right column (with the L1 branch). Instead of showing the frequency $\Omega$ itself, we show in the bottom row $C_{02}(\Omega; K)$ for the T branch and $C_0(\Omega; K)$ for the L1 and L2 branches. Only a few selected resonant branches are shown in the top and middle rows for clarity.}
\label{fig:double-gaussian}
\end{figure*}

All the collective modes have positive damping rates when the DLN distribution $g(\omega, \mathbf{v})$ has no zero crossing. If $g(\omega, \mathbf{v})$ has a zero crossing, however, some of the collective modes can acquire negative damping rates and become unstable. This is known as inverse Landau damping in plasma physics. We will first illustrate slow flavor instabilities in isotropic neutrino gases and then briefly discuss a few examples of fast flavor instabilities in anisotropic neutrino gases.

\subsection{Slow flavor instabilities}
\label{sec:sfi}

To illustrate slow flavor instabilities, we consider an isotropic neutrino gas with the following double-Gaussian DLN distribution, which has a single zero crossing:
\begin{align}
    g(\omega) = \phi(\omega; 0, \sigma) - \alpha\,\phi(\omega; \delta\omega, \sigma),
\end{align}
where $0< \alpha < 1$ and $|\delta\omega| \sim \sigma \ll \mu = 1$. We show the convolved DLN distributions $C_0(\varpi; K)$ and $C_{02}(\varpi; K)$ for two different double-Gaussian DLN distributions, one with negative $\delta\omega$ and the other with positive $\delta\omega$, in the middle and right panels of Fig.~\ref{fig:conv}, respectively. Unlike the case of the single-Gaussian DLN distribution, $C_0(\varpi; K)$ and $C_{02}(\varpi; K)$ now have zero crossings near $\varpi = \pm K$ depending on the sign of $\delta\omega$. As the frequency $\Omega$ of a metastable branch approaches the zero crossing, it will first become resonant with the neutrinos in the flipped (negative) section of the convolved DLN distribution, acquiring a positive growth rate. 

We show the T branch (with a few selected resonant transverse branches) in the left column of Fig.~\ref{fig:double-gaussian} for the DLN distribution with $\alpha=0.5$, $\delta\omega=-0.02$, and $\sigma=0.01$. At $K=0$, it reduces to a dipole mode which, according to Eq.~\eqref{eq:dipole}, has the frequency and growth rate
\begin{align}
    \Omega_\mathrm{d} \approx -\frac{1-\alpha}{3}
    \qquad\text{and}\qquad
    \Gamma_\mathrm{d} \approx -\pi\Omega^2_\mathrm{d} \frac{g(\Omega_\mathrm{d})}{1-\alpha},
\end{align}
respectively, where the factor $G_0 = 1-\alpha$ accounts for the change of the normalization of $g(\omega)$. It experiences inverse Landau damping because $g(\Omega_\mathrm{d}) < 0$,  but the growth rate is tiny since $|g(\Omega_\mathrm{d})| \ll 1$. With increasing $K$, this mode remains metastable until its frequency $\Omega$ enters (from the left) the effective support of $C_{02}(\varpi; K)$ that has a small negative section (middle panel of Fig.~\ref{fig:conv}). After that, it becomes a resonant mode with a growth rate increasing rapidly with $K$. Its growth rate then decreases and crosses zero where $C_{02}(\Omega; K) = 0$. It becomes a damped resonant mode at even larger $K$.

The behavior of the L2 branch shown in the middle column of Fig.~\ref{fig:double-gaussian} is qualitatively similar to that of the T branch. As discussed in Sec.~\ref{sec:longitudinal}, in the near-luminal regime where $|\Omega|\approx K$ is much larger than $|\delta\omega| \sim \sigma$ and $|\Gamma|$, its damping/growth rate is largely determined by the convolved DLN distribution $C_0(\varpi; K)$. In Fig.~\ref{fig:double-gaussian}, one can see that the inverse damping section of the L2 branch indeed approximately corresponds to the negative section of $C_0(\Omega; K)$. 

\begin{figure*}[tbp]
    \includegraphics[width=\textwidth]{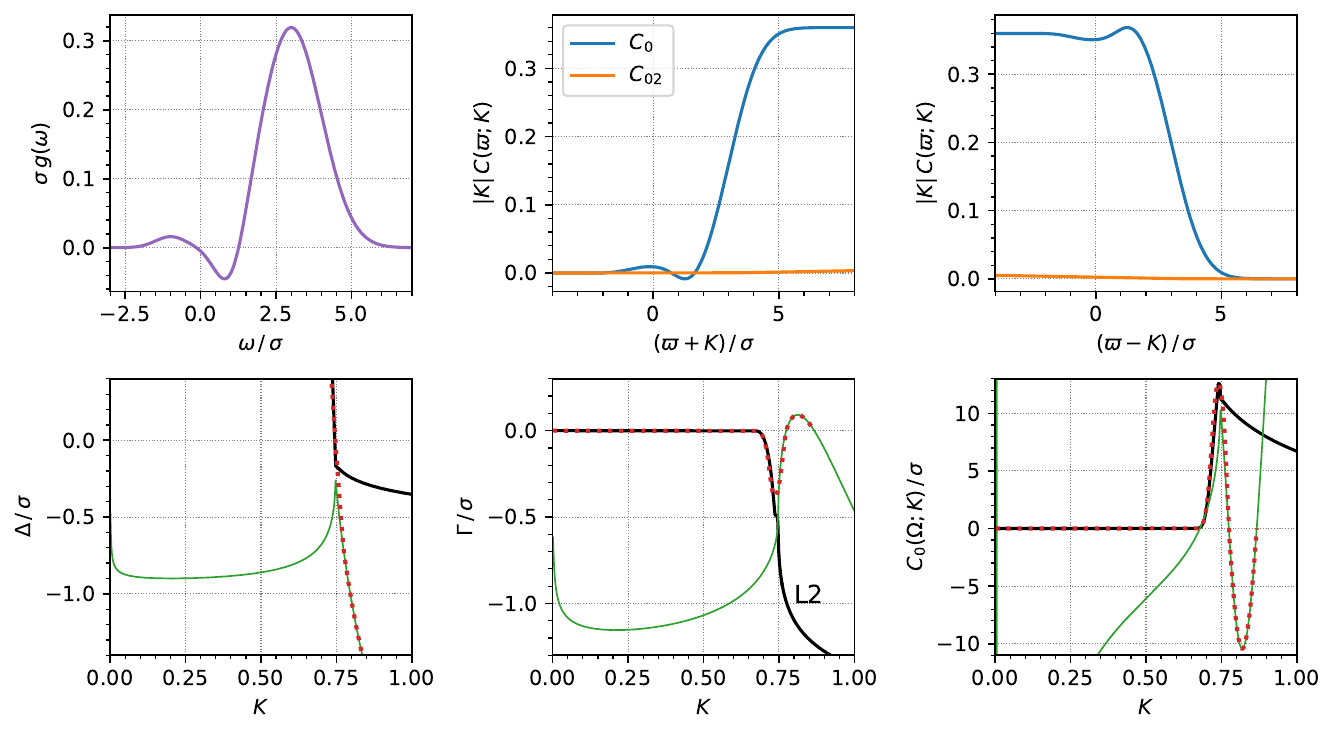}
    \caption{Top left: the triple-Gaussian DLN distribution $g(\omega) = 0.8\,\phi(\omega; 3\sigma, \sigma) - 0.1\,\phi(\omega; \sigma, \sigma/2) + 0.02\,\phi(\omega; -\sigma, \sigma/2)$ with $\sigma=0.001$. Top middle and top right: the left and right edges of the effective support of $C_0(\varpi; K)$ as a function of $\varpi$. Also shown is $C_{02}(\varpi; K)$ in the same region. Bottom row: distances from the light cone $\Delta$ (left), growth rates $\Gamma$ (middle), and $C_0(\Omega; K)$ (right) for the L2 branch (thick solid curve) and a resonant branch (thin solid curve) for the triple-Gaussian DLN distribution. The dotted curves are obtained by using the method described in Ref.~\cite{Fiorillo:2025zio} in which the integration is performed over velocity first to identify the Landau poles.}
    \label{fig:triple-gaussian}
\end{figure*}

Neither the T branch nor the L2 branch has any inverse damping section for positive $\delta\omega$. This is because, as $K$ increases from 0, their frequencies $\Omega$ enter the effective support of $C_0(\varpi; K)$ and $C_{02}(\varpi; K)$ from the left, which is always positive. (See the right panel of Fig.~\ref{fig:conv}.) The L1 branch, however, behaves differently from the T and L2 branches because its frequency $\Omega$ approaches the effective support of $C_0(\varpi; K)$ from the right instead of from the left. Therefore, we expect it to have an inverse damping section for a positive $\delta\omega$ but not for a negative $\delta\omega$. This is indeed the case, as shown in the right column of Fig.~\ref{fig:double-gaussian}.

All the branches in Fig.~\ref{fig:double-gaussian} that are gapless at $K=0$ remain damped for all $K$ because they are always resonant with significant portions of the convolved DLN distributions that are positive. This can change for more complicated DLN distributions. In Fig.~\ref{fig:triple-gaussian}, we examine the case of a triple-Gaussian DLN distribution with two zero crossings, similar to the example in Ref.~\cite{Fiorillo:2025kko}.
In this case, $C_{02}(\varpi; K)$ is highly suppressed in the near-luminal regime because of the factor $(1-u^2)$ in the integrand. In addition, $C_0(\varpi; K)$ has both a small positive section and a small negative section at the left edge of its effective support, and it has no negative section at the right edge. Naively, one may expect that the L2 branch would have an inverse damping section as its frequency $\Omega$ approaches the effective support of $C_0(\varpi; K)$ from the left. In this case, however, its frequency never enters the negative section of $C_0(\varpi; K)$ and, therefore, it does not experience inverse Landau damping. Interestingly, there is a resonant mode whose frequency $\Omega$ is in the negative section of $C_0(\varpi; K)$ even at small $K$. As $K$ increases, it first moves to the left (relative to the light cone) and enters the small positive section of $C_0(\varpi; K)$, and then moves back toward the negative section of $C_0(\varpi; K)$. Its damping rate decreases in this process and is determined by the value of $C_0(\Omega; K)$ when $\Gamma$ is small enough. The damping rate becomes negative when $\Omega$ is back in the negative section of $C_0(\varpi; K)$ and the mode becomes unstable. As $K$ increases further, it moves into the main positive section of $C_0(\varpi; K)$ and becomes a damped resonant mode again.  

In the bottom panels of Fig.~\ref{fig:triple-gaussian} we also plot, as dotted curves, the results obtained by using the method in Refs.~\cite{Fiorillo:2025zio,Fiorillo:2025kko}. One can see that this method mixes different branches of the collective modes and fails to track the DRs to large wave numbers. (See the discussion at the end of Sec.~\ref{sec:asymptotic}.)

\subsection{Fast flavor instabilities}
\label{sec:ffi}

\begin{figure*}[tbp]
    \includegraphics[width=\textwidth]{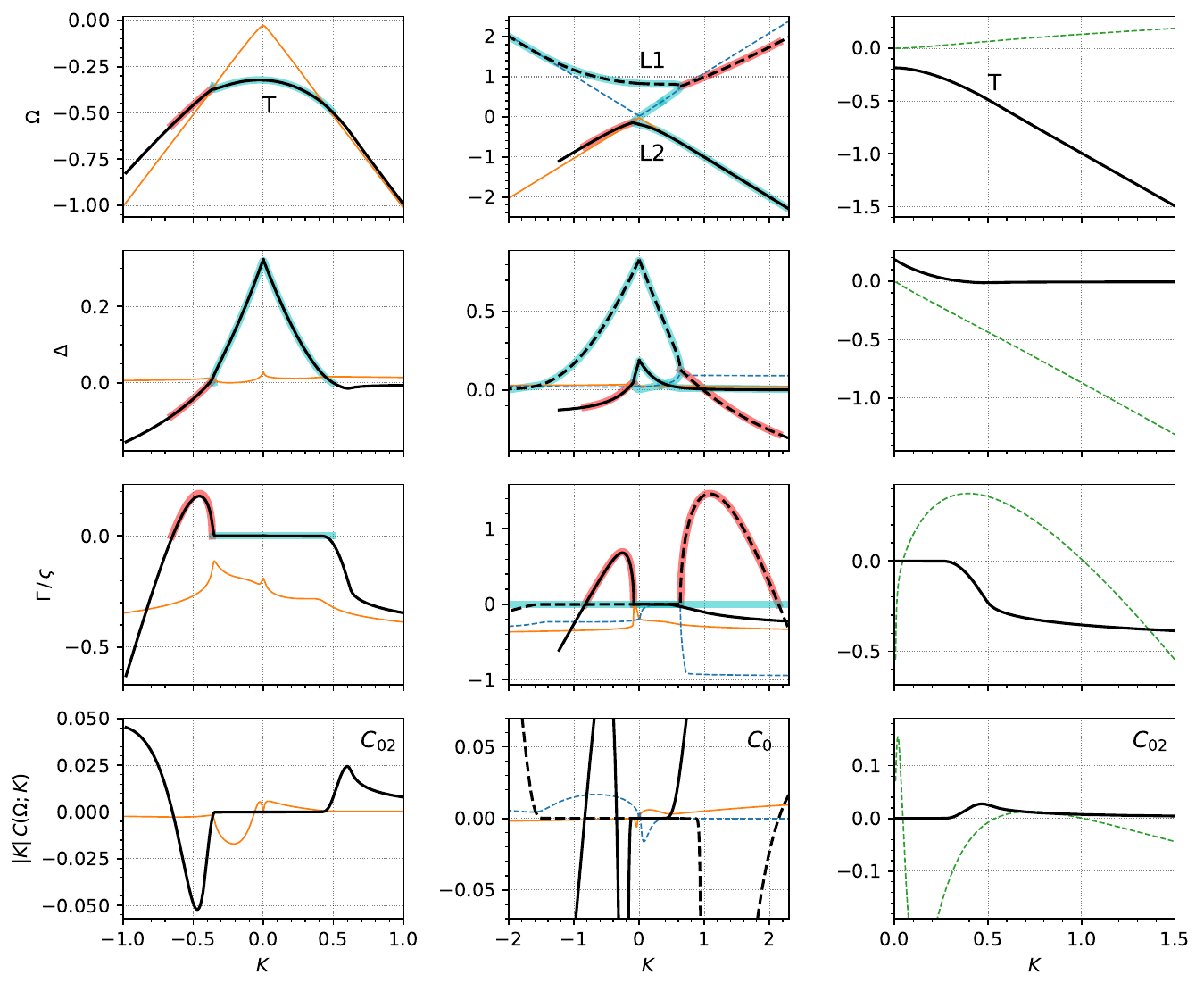}
    \caption{Similar to Fig.~\ref{fig:single-gaussian} but for the neutrino gases with anisotropic angular DLN distributions $\hat{g}(u) = 1 - 0.6\,\phi(u; u_0, \varsigma)$ with $\varsigma = 10 \sigma = 0.1$. Left and middle columns: some of the transverse and longitudinal branches for the DLN distribution with $u_0=1$. The semitransparent thick curves represent the unstable and stable branches in the fast limit $\sigma\to 0$. Right column: two transverse branches for the DLN distribution with $u_0=0$.}
    \label{fig:v-gaussian}
\end{figure*}

We now consider a neutrino gas with an axially symmetric DLN distribution of the form
\begin{align}
    g(\omega, \mathbf{v}) = \hat{g}(u) g(\omega),
\end{align}
where $u = \mathbf{v} \cdot \hat{\mathbf{z}}$ is the cosine of the angle between the neutrino velocity $\mathbf{v}$ and the $z$ axis, and $\hat{g}(u)$ and $g(\omega)$ are the angular and frequency DLN distributions, respectively. For simplicity, we use the single-Gaussian distribution in Eq.~\eqref{eq:single-gaussian} for $g(\omega)$ and angular DLN distributions of the form
\begin{align}
    \hat{g}(u) = 1 - \eta \phi(u; u_0, \varsigma),
    \label{eq:g-u}
\end{align}
where we assume $\varsigma$ to be much larger than the width $\sigma$ of the frequency DLN distribution $g(\omega)$.
We further assume that $\mathbf{K} = K \hat{\mathbf{z}}$, where $K$ can be either positive or negative. 
The convolved DLN distributions are then given by
\begin{align}
    C_n(\varpi; K) &= \int_{-1}^{1} u^n \hat{g}(u) 
    g(\varpi - K u)\,\frac{\mathrm{d}u}{2}
    \intertext{and}
    C_{02}(\varpi; K) &= \int_{-1}^{1} (1-u^2) \hat{g}(u) g(\varpi - K u)\,\frac{\mathrm{d}u}{2},
\end{align}
respectively. 
In the fast limit $\sigma \to 0$,
\begin{align}
    2 |K| C_0(\varpi; K)& \to 
    \hat{g}\left(\frac{\varpi}{K}\right) B\left(\frac{\varpi}{K}\right)
    \intertext{and}
    2 |K| C_{02}(\varpi; K)& \to 
    \left(1-\frac{\varpi^2}{K^2}\right) 
    \hat{g}\left(\frac{\varpi}{K}\right) B\left(\frac{\varpi}{K}\right),
\end{align}
where $B(x) = 1$ for $|x|<1$ and 0 otherwise. Therefore, a crossing of the angular DLN distribution $\hat{g}(u)$ implies a crossing of the convolved DLN distributions $C_0(\varpi; K)$ and $C_{02}(\varpi; K)$ in the fast limit. From the discussion in Sec.~\ref{sec:sfi}, we expect that, for a DLN distribution with a single angular crossing in the forward or backward direction, the metastable branches of the collective modes can become unstable when their frequencies $\Omega$ enter the negative (flipped) section of the convolved DLN distributions. This is indeed the case, as shown in the left and middle columns of Fig.~\ref{fig:v-gaussian}, where we show the DRs for the DLN distribution with $\eta=0.6$, $u_0=1$, and $\varsigma=10\sigma=0.1$.

In the left column of Fig.~\ref{fig:v-gaussian}, one observes that the T branch becomes a damped resonant branch as the wave number increases from 0 to $K\approx 0.5$. This is because its frequency approaches the effective support of $C_{02}(\varpi; K)$ from the left, which is always positive for $K>0$. For $K<0$, however, $C_{02}(\varpi; K)$ has a negative section at its left edge. As $K$ becomes increasingly negative, the T branch becomes an unstable resonant branch as its frequency enters the negative section of $C_{02}(\varpi; K)$. It then becomes a damped resonant branch when it enters the positive section of $C_{02}(\varpi; K)$ at even larger $|K|$.

In the middle column of Fig.~\ref{fig:v-gaussian}, we show the L1 and L2 branches, whose properties can be understood in terms of $C_0(\varpi; K)$ if $|\Delta|$ (and $\sigma$) is much smaller than $|K|$. Like the T branch, the L2 branch becomes a damped resonant branch as its wave number increases from 0 to $K\approx 0.5$, where its frequency enters the effective support of $C_0(\varpi; K)$ from the left, which is always positive if $K>0$. For $K<0$, $C_0(\varpi; K)$ has a negative section at its left edge. As $K$ increases in the negative direction, the L2 branch becomes unstable as $\Omega$ enters the negative section of $C_0(\varpi; K)$. Its behavior is not completely determined by $C_0(\varpi; K)$ at larger $|K|$ because it is no longer in the near-luminal regime and $\Gamma$ is appreciable. We were not able to track the L2 branch for $K\lesssim -1.5$ because of numerical difficulties. The L1 branch behaves similarly to the L2 branch except that it enters the effective support of $C_0(\varpi; K)$ from the right instead of the left. Therefore, it experiences inverse Landau damping in the positive $K$ direction instead of the negative $K$ direction. It becomes a damped resonant branch when $|K|$ is large enough in either the positive or negative direction.

For comparison, we also use the \texttt{FOWDR} Python package \cite{fowdr} to compute the stable and unstable branches in the fast limit $\sigma\to 0$, and the results are shown as the semitransparent thick curves in the left and middle columns of Fig.~\ref{fig:v-gaussian}. It is interesting to note that the stable longitudinal branch in the fast limit consists of segments of both the L1 and L2 branches and two damped resonant branches that nearly collide with the former two branches where they become unstable. The stable transverse branch in the fast limit also consists of a segment of the T branch and a small segment of a damped resonant branch that nearly collides with the T branch. 

As in the case of the slow flavor instabilities, the situation becomes more complicated when the angular DLN distribution has more than one crossing. In the right column of Fig.~\ref{fig:v-gaussian}, we show two transverse branches for an anisotropic neutrino gas with an angular DLN distribution of the form of Eq.~\eqref{eq:g-u} with $u_0=0$. This angular DLN distribution $\hat{g}(u)$ has two crossings near the center. In this case, the T branch becomes damped at $K\approx 0.3$ as its frequency enters the effective support of $C_{02}(\varpi; K)$, which is always positive at both edges. 
In fact, $C_{02}(\varpi; 0) \propto g(\varpi)=\phi(\varpi;0,\sigma)$. As $K$ increases, a dip develops in the middle of $C_{02}(\varpi; K)$, which eventually becomes negative. In the figure we show a resonant branch which has $\Omega\approx 0$ and becomes unstable when it enters the negative section of $C_{02}(\varpi; K)$. As $K$ continues to increase, the frequency $\Omega$ of this resonant branch moves to the right and enters the positive section of $C_{02}(\varpi; K)$. It becomes a damped resonant branch when $\Omega$ moves into the negative section of $C_{02}(\varpi; K)$ again, which indicates that $\partial_\Omega\mathcal{D}_\mathrm{T}(\Omega)$ has changed sign while $\Gamma$ is large. (See Eq.~\eqref{eq:Gamma-small}.) 

\section{Summary}
\label{sec:summary}

We have studied the collective flavor excitations in dense neutrino media using Landau's initial-value approach, in which the DRs of the damped modes are obtained by analytically continuing the self-consistency equation through a deformation of the integration contour. This is the same approach as that of the linear response theory developed in Refs.~\cite{Fiorillo:2024bzm,Fiorillo:2024uki,Fiorillo:2024pns,Fiorillo:2025ank,Fiorillo:2025zio}. The new element of this work is a more systematic treatment of DLN distributions that are continuous in the neutrino vacuum oscillation frequency $\omega$, and of how such distributions change the analytic structure of the solutions to the DR equation and the properties of the DR branches as functions of the wave number.

Some of our results follow from the theory without reference to a specific DLN distribution. A collective mode that is stable in the fast limit becomes metastable when the DLN distribution is nonzero at its resonance frequency, with a damping or growth rate given by Eq.~\eqref{eq:Gamma-small}. For a neutrino gas under typical supernova conditions, this damping rate is extremely small, which makes these modes practically stable. We have verified the Landau damping rate of the monopole mode in a homogeneous and isotropic neutrino gas by a direct numerical simulation of the flavor evolution (Fig.~\ref{fig:example}). 

Landau's formulation naturally leads to the convolved DLN distributions, which are the main conceptual tool of this work. We have defined them for axially symmetric DLN distributions, although they can be generalized to the non-axisymmetric case. At a given wave number $K$, $C_0(\varpi; K)$ is proportional to the net DLN of the neutrinos that are in kinematic resonance with a collective mode of frequency $\Omega=\varpi$. The corresponding quantity relevant to the transverse modes is $C_{02}(\varpi; K)$, which includes a velocity weighting imposed by the conservation of angular momentum in the emission and absorption of flavomons. The DR of the transverse modes is determined exactly by $C_{02}(\varpi; K)$ [Eq.~\eqref{eq:DR-T}], and that of the near-luminal longitudinal modes is determined approximately by $C_0(\varpi; K)$ (Appendix~\ref{sec:luminal}). The convolved DLN distributions thus play the same role at finite $K$ as the frequency DLN distribution $g(\omega)$ does for the homogeneous modes, and they reduce the stability analysis at each wave number to the examination of a one-dimensional function. A collective mode is metastable if its resonance region (approximately $\Omega - |\Gamma|$ to $\Omega + |\Gamma|$) lies outside the effective support of the relevant convolved DLN distribution, and it becomes resonant when its resonance region enters that support. For a mode with a small damping/growth rate $|\Gamma|$, whether it grows or decays is determined by the sign of the convolved DLN distribution at its frequency. This picture places slow and fast flavor instabilities on the same footing: both arise from the inverse Landau damping of a collective mode whose frequency falls into a flipped section of the convolved DLN distribution. In an isotropic gas, a crossing in $g(\omega)$ produces flipped sections near the edges of the light cone $\varpi\approx\pm K$. In the fast limit, the convolved DLN distributions become proportional to the angular DLN distribution $\hat{g}(\varpi/K)$, and an angular crossing becomes a crossing of the convolved DLN distributions.

For a DLN distribution with a single shallow crossing, the T, L1, and L2 branches of the collective modes are metastable at small $K$, and their frequencies approach the effective support of the convolved DLN distributions as $|K|$ increases.
By inspecting the edges of the convolved DLN distributions, one can therefore understand which branches may develop an inverse damping section and in which direction of $K$, without solving the DR equation. For DLN distributions with more than one crossing, a resonant branch, instead of the metastable branches, may become unstable (Fig.~\ref{fig:triple-gaussian} and the right column of Fig.~\ref{fig:v-gaussian}).

Some of the features of the DR branches in this work rely on the analytic properties of the DLN distribution assumed in Sec.~\ref{sec:theory}, namely that it is an entire function of $\omega$ that vanishes sufficiently fast as $\omega\to\pm\infty$, but not on its specific form. We have used combinations of Gaussian distributions in our examples because they allow some of the results to be obtained analytically. For the homogeneous modes, the DR equation can be solved in terms of the Faddeeva function, which shows that there are infinitely many resonant modes in addition to the metastable modes. These resonant modes correspond to the gapless modes in Ref.~\cite{Fiorillo:2025zio}. For the inhomogeneous modes, we have found numerically, and expect more generally, that the resonant branches do not terminate at specific wave numbers but extend across the entire range of wave numbers, that the metastable branches become resonant at large enough wave numbers, and that parts of the metastable or least damped resonant branches can become unstable when the convolved DLN distributions have flipped sections. Although these features may not all apply to DLN distributions that do not satisfy the assumptions in Sec.~\ref{sec:theory}, the insights developed in this work should still be useful for analyzing the collective modes in such cases.

Finally, we note that the general idea that the unstable modes and the Landau-damped modes are continuations of each other was already present in previous works on the pure fast modes (e.g., Ref.~\cite{Fiorillo:2024dik}). In the fast limit considered in Ref.~\cite{Fiorillo:2024dik} (see Fig.~5 therein), when there is no angular crossing ($\hat{g}(u)>0$ for all $u$), the two stable longitudinal branches continue for all wave numbers, and two separate, subluminal, resonantly damped branches terminate at the light cone. The stable branches join the damped branches at the light cone when the angular DLN distribution dips to 0 in the forward direction ($\hat{g}(1)=0$). As $\hat{g}(1)$ dips below 0, parts of the resonant branches become inversely damped, and a new stable section appears to join the two stable branches and form a single continuous stable branch. The fast limit, however, is an idealization of a neutrino gas with a small but finite spread in the vacuum oscillation frequency. With such a spread, the L1 and L2 branches remain superluminal in the absence of an angular crossing, but they become resonantly damped as they approach the light cone (Fig.~\ref{fig:single-gaussian}). There also exist both subluminal and superluminal resonantly damped branches near the light cone. When the angular DLN distribution develops a crossing, the L1 and L2 branches enter the light cone, where parts of them become inversely damped and nearly collide with two of the superluminal damped branches (Fig.~\ref{fig:v-gaussian}). The sections of these two superluminal branches that lie between the L1 and L2 branches, in both $K$ and $\Omega$, have damping rates much smaller than the rest of the same branches, and these damping rates vanish as the spread in vacuum frequency approaches 0. These sections correspond to the new stable section that appears abruptly in the fast limit. We therefore expect the abrupt appearance of the S shape to be the fast-limit counterpart of a continuous change of the DR branches when the spread in vacuum frequency is finite.

\begin{acknowledgments}
We thank D.~F.~G.~Fiorillo for helpful discussions. This work was supported in part by the US DOE NP grant No.\ DE-SC0017803 at UNM.
\end{acknowledgments}

\section*{Data Availability}
The data that support the findings of this article, together with the Jupyter notebooks used to generate the figures, are openly available~\cite{Kost:data}.

\appendix
\section{Damping/Growth Rates of Longitudinal Modes in the Near-Luminal Regime}
\label{sec:luminal}

In Sec.~\ref{sec:longitudinal}, we argued, based on conservation of angular momentum, that the properties of the longitudinal modes in the near-luminal regime are determined by the convolved DLN distribution $C_0(\varpi; K)$. Here, we obtain the growth rate $\Gamma$ in this regime, assuming it is small. Suppose that the longitudinal modes have the following DR equation in this regime:
\begin{align}
    0 = \mathcal{D}_\mathrm{L} \approx a Z_0(\Omega + \mathrm{i}\Gamma, K) - b,
\end{align}
where $a$ and $b$ are real constants, and
\begin{align}
    Z_0(\Omega + \mathrm{i}\Gamma, K) = \int_\mathcal{L} \frac{C_0(\varpi; K) \,\mathrm{d}\varpi}{\Omega + \mathrm{i} \Gamma - \varpi}.
\end{align}
Using Eq.~\eqref{eq:Gamma-small}, we find
\begin{align}
    \Gamma & \approx -\frac{\mathrm{Im}[Z_0(\Omega, K)]}{\partial_\Omega \mathrm{Re}[Z_0(\Omega, K)]} 
\end{align}
for a weakly damped or growing mode of wave number $K$ and frequency $\Omega$.

Let us consider a neutrino gas with an axially symmetric DLN distribution $g(\omega, u)$ for which
\begin{align}
    C_0(\varpi; K) = \int_{-1}^{1} g(\varpi - K u, u)\,\frac{\mathrm{d}u}{2}.
\end{align}
In the limit where $|\Omega - K|$ and $\sigma$ (the spread of the frequency DLN distribution) are both much smaller than $|K|$, one has
\begin{subequations}
 \begin{align}
    \mathrm{Im}[Z_0(\Omega, K)] 
    & = -\pi C_0(\Omega; K) 
    \\
    &= -\pi \int_{-1}^{1} g(\Omega - K u, u)\,\frac{\mathrm{d}u}{2} 
    \\
    &\approx -\frac{\pi}{2K} \int_{\Omega - K}^{\Omega + K} g(\omega, 1)\,\mathrm{d}\omega.
\end{align}   
\end{subequations}
Meanwhile,
\begin{subequations}
\begin{align}
    \mathrm{Re}[Z_0(\Omega, K)] 
    &= \mathrm{PV} \int_{-\infty}^{\infty} \mathrm{d}\omega
    \int_{-1}^1 \frac{g(\omega, u)}{\Omega - \omega - K u} \,\frac{\mathrm{d}u}{2} 
    \\
    &\approx \frac{-1}{2K} \mathrm{PV} \int_{-\infty}^{\infty} \mathrm{d}\omega\,
    g(\omega, 1)\ln(\Omega - \omega - K),
\end{align}  
\end{subequations}
where we have kept the most divergent term in the limit where $|\Omega - \omega - K|\approx 0$.
Therefore,
\begin{align}
    \partial_\Omega \mathrm{Re}[Z_0(\Omega, K)] 
    &\approx \frac{-1}{2K} \mathrm{PV} \int_{-\infty}^{\infty} \frac{g(\omega, 1)\,\mathrm{d}\omega}{\Omega - \omega - K},
\end{align}
and
\begin{align}
    \Gamma & \approx  \frac{\displaystyle -\pi \int_{\Omega - K}^{\Omega + K} g(\omega, 1)\,\mathrm{d}\omega}
    {\displaystyle\mathrm{PV} \int_{-\infty}^{\infty} \frac{g(\omega, 1) \,\mathrm{d}\omega}{\Omega - \omega - K}}.
    \label{eq:Gamma-lum}
\end{align}
This result agrees with the approximation in Refs.~\cite{Fiorillo:2025zio,Fiorillo:2025kko} for the damping/growth rate of the near-luminal longitudinal modes when the integration limit $\Omega + K$ is replaced by $\pm\infty$ depending on whether $K$ is positive or negative. 
Equation~\eqref{eq:Gamma-lum} also applies to the other near-luminal regime with $|\Omega + K| \ll |K|$ except with the replacements $K\to -K$ and $g(\omega, 1)\to g(\omega, -1)$.

\bibliography{damping}

\end{document}